\documentclass[12pt]{article}
\usepackage{epsf}
\usepackage{graphicx}

\begin{document}
\title
{Low-energy quantum gravity: new challenges for an experiment and
observation}
\author
{Michael A. Ivanov \\
Physics Dept.,\\
Belarus State University of Informatics and Radioelectronics, \\
6 P. Brovka Street,  BY 220027, Minsk, Republic of Belarus.\\
E-mail: michai@mail.by.}

\maketitle

\begin{abstract}
Some new challenges for an experiment and observation, which are
consequences of the model of low-energy quantum gravity by the
author, are considered here. In particular, the property of
asymptotic freedom of this model leads to the unexpected
consequence: if a black hole arises due to a collapse of a matter
with some characteristic mass of particles, its full mass should
be restricted from the bottom. For usual baryonic matter, this
limit of mass is of the order $10^{7}M_{\odot}$.
\end{abstract}

During a few last decades, a verification of general relativity
was almost a synonym for any experimental work in gravity. With
enviable constancy this theory was recognized again and again as a
favorite one among others, without any inconsistences with
observations. But any theory should have its own borders of
applicability. I think that we saw them for general relativity
only in 1998, when Anderson's team reports about the Pioneer
anomaly \cite{1}; this effect is obviously not embedded in a frame
of general relativity. In 2002, Nesvizhevsky's team reported about
discovery of quantum states of ultra-cold neutrons in the Earth's
gravitational field \cite{4}. Observed energies of levels have the
order of $10^{-12}$ eV. It means that energies of irradiated
gravitons are of 40 orders lesser than the Planck energy. It is an
absolutely unexpected scale for quantum gravity, but many prefer
to think that this result is not connected with quantum gravity!
The discovery of dimming of remote supernovae 1a in 1998
\cite{2,3} led to an introduction into physics of some new
component - "dark energy", which is unknown from any laboratory
experiment. It is very intriguing for me that a majority of people
trust in this explanation of the observational peculiarity of the
Hubble diagram; for me, it is simpler to doubt in the present
cosmological paradigm and in applicability of general relativity
for very big distances and time intervals. Another problem, the
one of missing mass or "dark matter" for galaxies, is much longer
standing. And there is not any warranty, too, that general
relativity is true on the galactic scale.
\par I would like to pay your attention here on new challenges for an
experiment and observation which are consequences of my model of
low-energy quantum gravity \cite{500}. In this model, quantum
gravity is considered as a very-low-energy phenomenon: the average
graviton energy is of the order of $10^{-3}$ eV. There are the
following main problems.
\par 1. {\it A verification of the redshift mechanism of this
model.} The redshift is caused by forehead collisions with
gravitons in this model. To verify this conjecture, the laser
experiment may be performed on the Earth \cite{112,500}. A price
of this question is very high: it would be possible to check
indirectly and the conjecture about an expansion of the Universe.
\par 2. {\it The Pioneer anomaly \cite{1}.} In the model, this
anomaly is analogical to the redshift for photons. There exist
plans of further investigation of this effect \cite{113}, and I
would like only to say that if my explanation is true then some
peculiarities should take place: the best parameter of the
anomalous acceleration should be the angle between a velocity of
the probe and its radius-vector (it means that the effect may
change its sign); a periodic contribution should exist due to an
anomalous acceleration of the Earth \cite{114}.
\par This deceleration of massive bodies by the graviton
background may lead to an additional relative acceleration of
bodies. Perhaps, namely the fact serves as a cause of successes of
MOND by M. Milgrom in explanation of flat rotation curves of
galaxies \cite{99}. In MOND, when a body acceleration gets the
threshold value of $\sim Hc,$ one introduces by hand the growth of
interaction; but this value characterizes the Pioneer anomaly in
my model. Another possible origin of flat rotation curves in this
model may be the screening of internal parts of a galaxy with its
external parts, that will lead to a relative magnification of
attraction of a periphery to the center.
\par 3. {\it A multivalued character of the Hubble diagram.} The
Hubble diagram is a multivalued function in this model \cite{115}.
It is difficult to verify this prediction, because GRBs are not
good cosmological candles, and supernovae 1a are not observable by
big enough redshifts.
\par 4. {\it The problem of existence of black holes.} The accepted
mechanism of gravity in the model \cite{500} leads to the
consequence that a black hole should have an essentially bigger
gravitational mass than an inertial one (approximately of 1000
times). There are the two variants: a) the equivalence principle
is valid, then black holes cannot exist in the nature (in this
case, super massive compact objects at centers of galaxies should
have another nature); b) the equivalence principle is not valid
for black holes which exist in the nature. In the second case,
black holes should aim to the dynamical center of a galaxy with a
huge acceleration due to the difference of gravitational and
inertial masses. The objects known as black holes correspond to
this scenario.
\par Additionally, the property of asymptotic freedom of this
model \cite{116} leads to the unexpected consequence: if a black
hole arises due to a collapse of a matter with some characteristic
mass of particles, its full mass should be restricted from the
bottom. For example, in a case of collapsing usual baryonic matter
one may accept that a particle mass is equal to the proton mass
$m_{p}.$ Big deviations from general relativity should take place
by the minimum radius of the object: $r_{min} \sim
<\sigma>^{1/2}N^{1/3},$ where $<\sigma>$ is an average
cross-section of an interaction of a particle with a graviton, $N$
is a full number of particles. We can compute the ratio
$r_{g}/r_{min},$ where $r_{g}=2Gm/c^{2}$ is a gravitational radius
of the object: $$r_{g}/r_{min} \sim (m/m_{0})^{2/3},$$ where
$m_{0} = m_{p}(<\sigma>^{1/2}/r_{gp})^{3/2}$, and $r_{gp}$ is a
formally introduced gravitational radius of proton. The rough
estimate for $m_{0}$ is: $m_{0} \sim 10^{7}M_{\odot}$. It is
necessary to have $r_{g}/r_{min} > 1$, or $m/m_{0} > 1$.
\par For another mass of particles of collapsing object, it is
easy to re-calculate this bottom limit of the mass; because
$m_{0}\sim m_{p}^{1/4}$, we shall have by some new mass of
particles $m':$ $m_{0}(m')=m_{0}(m_{p})(m'/m_{p})^{1/4}$.
\par 5. {\it Gravitational asymptotic freedom.} An unalienable
property of this model is asymptotic freedom at small distances
\cite{116}. The range of non-universal transition to asymptotic
freedom for protons is between $10^{-11} - 10^{-13}$ meter, and
for electrons it is between $10^{-13} - 10^{-15}$ meter. Big
efforts were undertaken recently to detect micron-scale deviations
from Newtonian gravity (for example, see \cite{117,118}), but this
new needed range is very far from the investigated limit.
\par 6. {\it Galaxy/quasar number counts}. Given only the
luminosity distance and a geometrical one as functions of a
redshift in this model, theoretical predictions for galaxy/quasar
number counts may be found \cite{119}. But the result depends on a
chosen kind of the luminosity function and a theoretical model of
quasar activity.
\par 6. {\it A violation of the postulate about constancy of the
velocity of light}. Due to a non-zero duration of an interaction
of photons with gravitons, this postulate should be violated in
the considered model if we consider very big distances. This
theoretical problem is now open. If by attempts to build a model
of quantum gravity starting from general relativity the small
parameter to describe violations of the postulate is the ratio
$E/E_{Pl}$, where $E$ is an energy of a photon, and $E_{Pl}$ is
the Planck energy \cite{120}, in this model we should consider the
ratio $\varepsilon/E$ as such the small parameter, where
$\varepsilon$ is an energy of a graviton. A duration of one act of
interaction would be estimated on a base of the uncertainties
relation, and one might find a photon delay on its way using the
lows of conservation of an energy and a momentum. I think that a
dispersion of time-in-flight should depend on the photon energy
(it should rise when $E$ decreases). Any efforts to observe or to
limit the Lorenz violation (similar to \cite{121}) are very useful
to clarify this question.
\par 7. {\it A connection of the two backgrounds}. The graviton
background should interact with the cosmic microwave one in this
model. Perhaps, one of consequences of this interaction would be
observed due to an existence of advanced technics and devices for
microwave radiation measurements: any source of gravitational
waves of general relativity should modulate the first background
that will lead to the similar {\it temporal} modulation of CMB on
expected frequencies of the gravitational waves. The important
characteristic feature of namely this connection exists: when the
modulated signal arrives to an observer from the source direction,
this modulation should appear from the opposite direction, too.
This proposal does not take into account a possibility of fast
relaxation of any disturbance in such the dynamical substance as
the background of super-strong interacting gravitons of this
model. The theoretical problem of dissipation of energy of
gravitational waves on their way in such the graviton background
is open, too.
\par In this model, one does not need any dark energy or an expansion
of the Universe to explain main observational facts. But there are
many open problems, some of them are discussed above, which should
be further investigated.

\end{document}